# PERFORMANCE ANALYSIS OF SYMMETRIC KEY CIPHERS IN LINEAR AND GRID BASED SENSOR NETWORKS

Kaushal Shah[1] and Devesh C. Jinwala[2]

[1]Department of Computer Engineering, S. V. National Institute of Technology, Surat, India. shah.kaushal.a@gmail.com
[2] Department of Computer Engineering, S. V. National Institute of Technology, Surat, India. dcjinwala@gmail.com

## ABSTRACT

*The linear and grid based Wireless Sensor Networks (WSN) are formed by applications where objects being monitored are either placed in linear or grid based form. E.g. monitoring oil, water or gas pipelines; perimeter surveillance; monitoring traffic level of city streets, goods warehouse monitoring. The security of data is a critical issue for all such applications and as the devices used for the monitoring purpose have several resource constraints (bandwidth, storage capacity, battery life); it is significant to have a lightweight security solution. Therefore, we consider symmetric key based solutions proposed in the literature as asymmetric based solutions require more computation, energy and storage of keys. We analyse the symmetric ciphers with respect to the performance parameters: RAM, ROM consumption and number of CPU cycles. We perform this simulation analysis in Contiki Cooja by considering an example scenario on two different motes namely: Sky and Z1. The aim of this analysis is to come up with the best suited symmetric key based cipher for the linear and grid based WSN.*

## 1. INTRODUCTION

The Wireless Sensor Networks (WSNs) are considered to be deployed in random or tree based fashion. However, there are applications of WSN that form specific topology like linear or grid based WSN. The applications where objects being monitored are distributed in either linear or square grid inherently form linear and grid based WSN. Examples of the same are:

i. Monitoring the traffic level of city streets.

ii. Monitoring pipelines carrying oil, water or gas.

iii. Monitoring goods in a warehouse.

iv. Perimeter surveillance.

The typical examples of nodes forming a linear network and square grid are as shown in Fig. 1 and 2 respectively. As shown in Fig. 1, there are 7 nodes deployed in a linear fashion and each node has a communication range of 2. Therefore, such network is known as (7, 2) linear network.

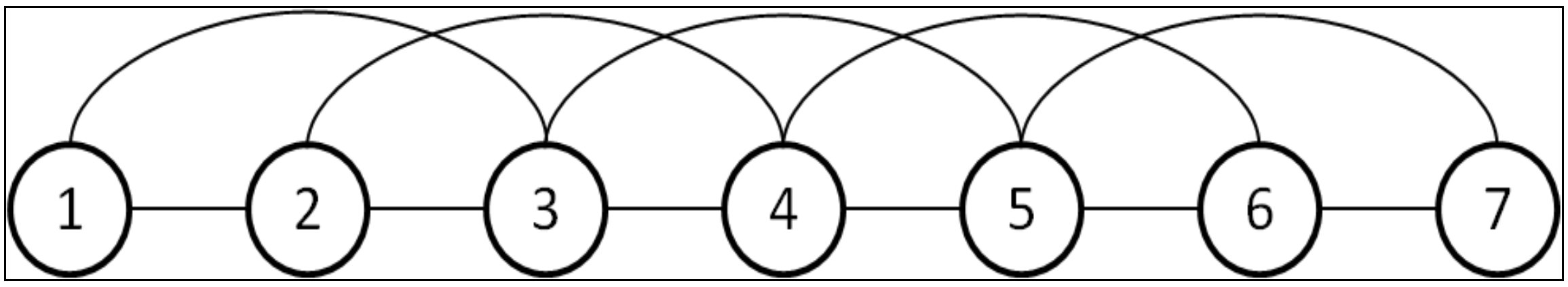

Fig. 1. A (7, 2)-linear sensor network.

As shown in Fig. 2, there are 50 houses of a colony deployed in a square grid manner. Here, each house is considered to have a device that is used for monitoring the energy consumption of the house. These devices pass the aggregated data in a hop by hop manner to the aggregator node, which send the data to the BS. Based on the considered application, there can be multiple aggregator nodes. The same deployment of devices can be considered for the applications of grid based networks. When we consider a single row of houses, it forms a linear network of 10 houses. Therefore, the grid based WSN are formed through the combination of linear networks (when 5 rows are considered, it forms grid based WSN). As the data are passed in hop by hop manner for getting the advantage of aggregation as discussed in [1], the security of the same is a critical issue. The intermediate nodes can alter or passively monitor the data and use it for their own advantage as discussed in [2, 3]. Therefore, the data are required to be encrypted before passing to the next node and the same is discussed in [4-6].

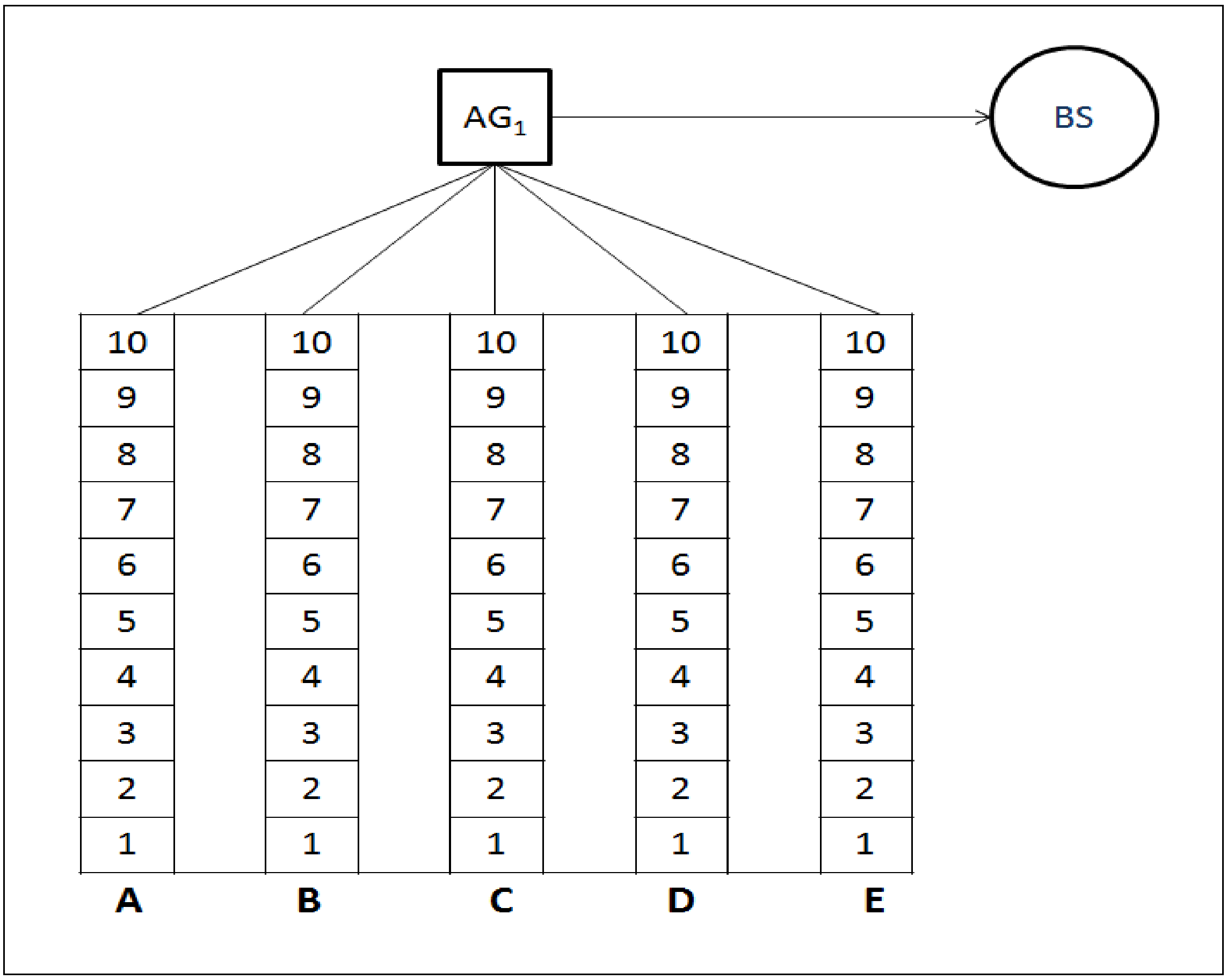

Fig. 2. Considered scenario of colony forming grid based network.

The lightweight schemes for the same are discussed in [7, 8]. However, these schemes do not take the deployment knowledge of sensor nodes under consideration unlike the secure data aggregation scheme for linear WSN proposed in [9]. For protecting the privacy of data, there are schemes proposed in the literature [10-13]. The scheme discussed in [10] is based on the idea of pseudonym changing. The idea of anonymous communication is presented in [11]. The key establishment scheme for the networks monitoring roads is discussed in [14]. The energy efficient communication protocol based on the idea of clustering in road networks is presented in [15]. Using the public key cryptography without using the certificates, the authentication scheme is discussed in [16]. In this paper, we perform the simulation analysis to find out the number of CPU cycles and memory consumption required in providing data security for linear and grid WSN.

The devices (sensor nodes) used in the considered applications are constrained with regard to the computation capabilities, battery power, storage, bandwidth [17, 18]. Therefore, we perform an analysis on the symmetric key based ciphers (as asymmetric key based ciphers require more storage and computation) and analyse the same for the considered example scenario. The performance parameters used in the simulation analysis are: RAM, ROM consumption and number of CPU cycles. The keys are pre-distributed as proposed in [19]. Such analysis is already discussed in the literature [20-22]. However, they have not considered the knowledge of deployment of nodes and the specific topology formed by the application. The main objective of our work is to come up with the best suited cipher for the considered scenario of linear and grid based WSN. Moreover, we perform our analysis on two different motes: sky and z1. The z1 motes have more capability regarding the RAM and the ash memory and the analysis results shows that they require lesser number of CPU cycles as compared to sky motes. In addition, the comparison of two operating systems contiki cooja and tinyOS for sensor nodes is discussed in the paper. We perform a detailed analysis for one of the most significant cryptographic primitives of WSNs: Symmetric Key Block Cipher. We consider the performance parameters, storage and energy for a set of candidate lightweight ciphers. Analysing the performance of the symmetric key based ciphers with respect to the performance parameters: number of CPU cycles and RAM,ROM in contiki cooja, is the major contribution of this paper.

### 1.1 Organization of the Paper

The rest of the paper is organized as follows: In Section 2, brief introduction about the examined symmetric key based ciphers is discussed. In Section 3, we look at the simulation setup and the methodology for the evaluation of the ciphers. Moreover, the comparison of the operating systems contiki cooja and tinyOS is discussed in this section. In Section 4, we show the simulation results and discuss the same. We summarize our work with conclusions in Section 5.

## 2. THE SYMMETRIC KEY BASED CIPHERS: EXAMINED

In this section, we discuss block ciphers that are lightweight in nature. Different ciphers are based on different structures like Substitution Permutation Network (SPN), Feistel or Lai-Massey. AES [23]; KLEIN [24]; LED [25]; PRESENT [26] are based on SPN structure. HIGHT [27]; LBlock [28]; MIBS [29]; PICCOLO [30]; SEA [31]; SIMON [32]; TWINE [33] are based on Feistel structure. SPECK [32] is based on ARX (Add-Rotate-Xor) structure. IDEA [34] is based on Lai-Massey structure. We select these ciphers for the analysis as the applications we considered require lightweight solution and with the analysis of the paper we come up with the best suited cipher from the considered lightweight ciphers. We provide an overview of the examined ciphers and the attacks that are possible on each of them. We do not go in the designing details of the ciphers as our main focus is to analyse the performance in terms of CPU cycles and RAM, ROM consumption.

## 2.1. Substitution Permutation Network (SPN) Structure and Related Ciphers

SPN structure [35] takes a plaintext block and a key as inputs, and performs exchanging "rounds" of substitution (S) and permutation boxes (P-boxes) respectively to deliver the ciphertext block. An S-box substitutes a block of bits given as input by another block of bits as the output. This substitution must guarantee invertibility. A P-box is a permutation of all the bits: it takes input from the outputs of all the S-boxes of previous round, applies permutation of bits, and augments them into the S-boxes of the following round. The key is combined using some group operation like XOR at each round. The S-boxes and P-boxes transformations are efficient to perform in device (like sensors), E.g. exclusive or (XOR) and bitwise rotation.

The following ciphers are based on SPN structure:

***Advanced Encryption Standard (AES).*** We analyse two different implementations of Advanced Encryption Standard (AES) ciphers. One is publicly available and other is designed by contiki cooja developers. Contiki has LLSec (Link Layer Security) layer. This layer is hardware independent, as it uses generic AES driver API instead of directly accessing the hardware. There are multiple AES drivers implemented in Contiki - software-only version and a couple of hardware accelerated ones, including for CC2420 (the radio chip on Sky mote). Authors of [36] show a possible attack on AES, known as biclique cryptanalysis. It uses the concept of exhaustive search on the key with an improvement by linking the keys through key schedule. This attack takes a time complexity of $2^{126.2}$ AES encryptions on the data amount $2^{88}$. The other possible attack is meet-in-the-middle [37] that takes less than $2^{100}$ data/time/memory complexity.

***KLEIN.*** We analyse two different implementations of this cipher: KLEIN64 and KLEIN96. Both the implementations take 64 bits block size. Key lengths are 80 and 96 bits respectively. The number of rounds can either be 12, 16, or 20. The possible attack on this cipher is chosen plaintext key recovery as discussed in [38].

***LED.*** We analyse two different implementations of this cipher: LED64 and LED128. Both the implementations take 64 bits block size. Key lengths are 64 and 128 bits respectively. The number of rounds is 32 and 48 respectively. This cipher does not use key schedule and this is the main difference from other ciphers. The XORing of key is done after every four rounds instead of key schedule. The number of rounds of this cipher is more as compared to other ciphers for compensating the key schedule. The differential cryptanalysis results on this cipher are discussed in [39]. The attacks on LED64 can be reduced to 12 and 16 rounds is described by the authors. The other possible attack is meet-in-the-middle as discussed in [40]. The complexity of the attack on 8 rounds of LED64 and 16 rounds of LED128 is lesser as compared to exhaustive key search.

***PRESENT.*** It is the most popular cipher among all lightweight block ciphers. We analyse two different implementations of this cipher: PRESENT Size and PRESENT Speed. Both implementations take 64 bits block size. Key lengths are either 80 or 128 bits with number of rounds as 31. There many cryptanalysis results as discussed in [41-43]. Authors in [44] discuss about two bicliques possible on two implementation of PRESENT.

## 2.2. Feistel Structure and Related Ciphers

Feistel structure [45] takes plaintext block as input and divides it into two halves, L (left) and R (right). R half is given as input to a feistel function along with the round key. It is also used as an L half for the next round. Output of the feistel function is XORed with L half and used as an R half for the next round. The same process is repeated till last round. The advantage with this structure is, just by reversing the key schedule decryption can be done.

The following ciphers are based on feistel structure (or a modified feistel structure):

***HIGHT.*** This cipher uses block size of 64 and key length of 128 bits. It uses 32 rounds and sometimes uses modular addition instead of XOR operation. The biclique attack against HIGHT is proposed in [46]. Moreover, the differential cryptanalysis attack is described in [47].

***LBLOCK.*** This cipher uses block size of 64 and key length of 80 bits. The number of rounds is 32 and the usage of 8 S-boxes and permutation of 4 bits are applied. The biclique attack against LBLOCK is proposed in [48]. The authors also discuss the prevention of this attack with the help of modified key schedule algorithm.

***MIBS.*** We analyse two different implementations of this cipher: MIBS64 and MIBS80. Both implementations take 64 bits block size. Key lengths are 64 and 80 bits respectively. The linear attacks on MIBS are discussed in [49]. The authors show the differential cryptanalysis on 14 rounds, ciphertext only attacks on 13 rounds and an impossible differential attack on 12 rounds of MIBS.

***PICCOLO.*** We analyse two different implementations of this cipher: PICCOLO80 and PICCOLO128. Both implementations take 64 bits block size. Key lengths are 80 and 128 bits respectively. It uses two feistel functions. It requires less than 1000 gates when implemented on hardware. Authors of [44] discuss the biclique attacks on both the implementations of PICCOLO.

**SEA.** This cipher uses *n* bits block size. The value of *n* can be 48, 96, or 144 bits. It uses a two branch feistel structure as modified feistel structure. The security analysis is discussed in [31].

***SIMON.*** This cipher uses different block sizes like 32, 48, 64, 96, 128 bits. It is based on a balanced feistel network. The key size can be 64, 72, 96, 128, 144, 192, 256 bits. It is optimized for the hardware implementations. The differential cryptanalysis is possible on this cipher as discussed in [50, 51].

***TWINE.*** We analyse two different implementations of this cipher: TWINE80 and TWINE128. Both the implementations take 64 bits block size. Key lengths are 80 and 128 bits respectively. The number of rounds is 36 in both the implementations. The feistel function uses a single Sbox and subkey addition. This function is repeated 8 times in each round. Two biclique attacks on two implementations are discussed in [52].

## 2.3. Add-Rotate-Xor Structure and Related Cipher

This structure involves 3 operations:

1. Modular Addition
2. Rotation with fixed rotation amounts
3. XOR

These ARX operations are immune to timing attacks because they run in defined constant time. As these operations are fast and cheap in hardware and software, the ciphers based on ARX operations are popular.

***SPECK.*** This cipher uses different block sizes like 32, 48, 64, 96, 128 bits. It is based on an Add-Rotate-Xor (ARX) structure. The key size can be 64, 72, 96, 128, 144, 192, 256 bits. It is optimized for the software implementations. The differential cryptanalysis is possible on this cipher as discussed in [50, 51].

## 2.4. Lai-Massey Structure and Related Cipher

As shown in Fig. 3, Lai-Massey Structure [53] divides the plaintext in two equal halves $L_0$ and $R_0$ as input. Two round functions are used; H and F. Keys are used with function F. The output of function H is given as input to function F.

$$(L_{i+1}{}', R_{i+1}{}') = H(L_i{}' + T_i, R_i{}' + T_i)$$
$$where\, T_i = F(L_i{}' - R_i{}', K_i)\; and\; (L_0{}', R_0{}') = H(L_0, R_0)$$

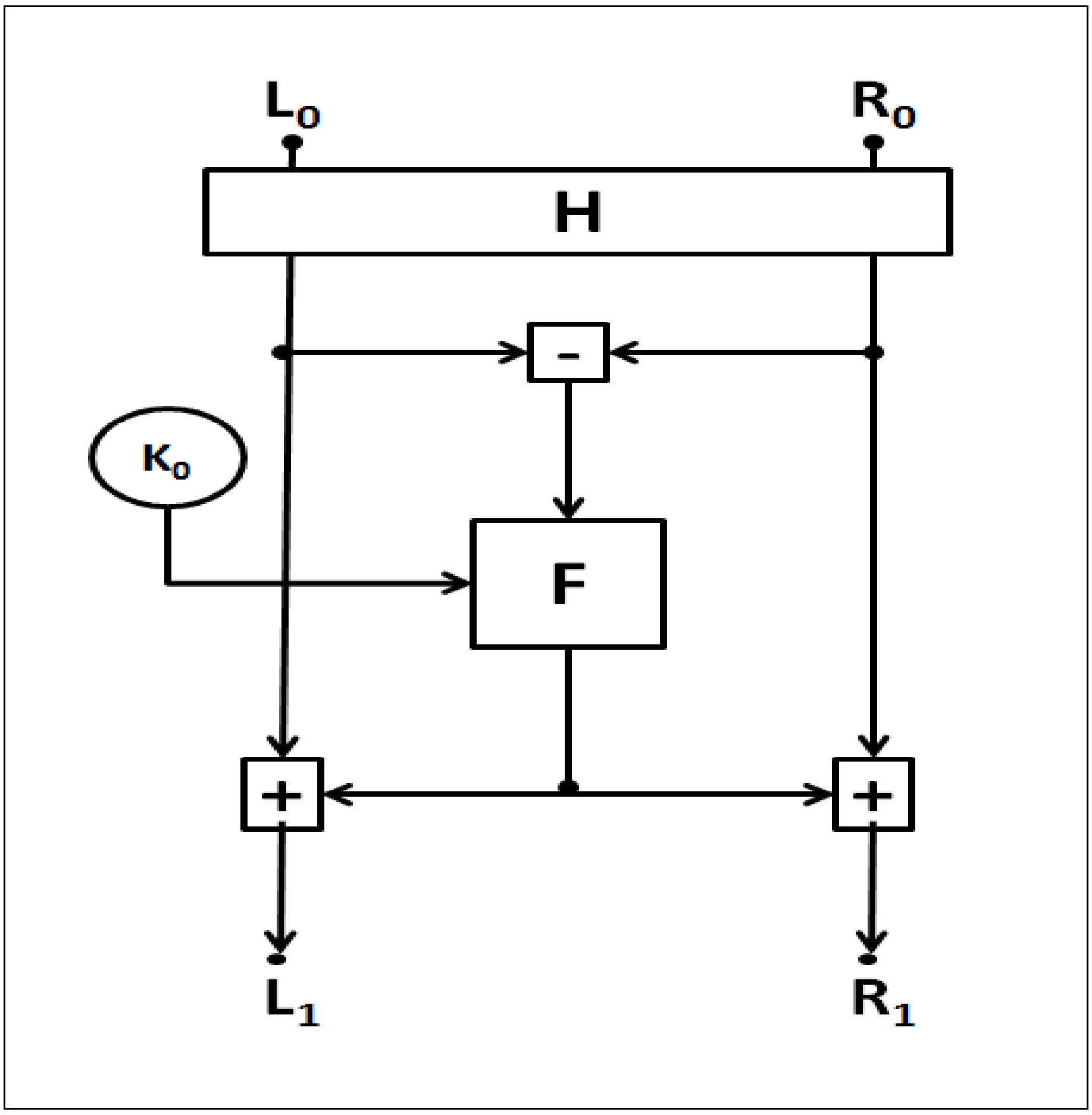

Fig. 3. Lai-Massey Structure

***IDEA.*** This cipher uses 64 bits block size and 128 bits key. It is composed of 8.5 rounds. A final round that is "half round", comes after eight rounds and used for the output transformation (the swap of the centre two values counterbalances the swap toward the end of the last round, so that there is no net swap). It is included in the package PGP (Pretty Good Privacy). There is six rounds attack that exploits key schedule of IDEA with linear cryptanalysis as discussed in [54]. The biclique framework is used by the authors of [55] to speed up the key recovery.

The summary of all the examined symmetric key based block ciphers with regard to the structure they are built on, block size, key size, and the attacks that are possible on each of them is shown in Table 1. From all the examined ciphers, SIMON and SPECK are considered to be the block ciphers for IoT (Internet of Things) environment as discussed in [56].

Table 1. Summary of Block Ciphers Examined.

| **Sr. No.** | **Cipher** | **Reference** | **Block Size (bits)** | **Key Size (bits)** | **Structure** | **Possible Attacks** |
|---|---|---|---|---|---|---|
| 1 | AES | [23] | 128 | 128 | SPN | - Biclique cryptanalysis [36]<br>- Meet-in-the-middle [37] |
| 2 | HIGHT | [27] | 64 | 128 | Fiestel | - Differential cryptanalysis [47] |
| 3 | IDEA | [34] | 64 | 128 | Lai-Massey | - Linear cryptanalysis [54]<br>- Biclique [55] |
| 4 | KLEIN | [24] | 64 | 64,96 | SPN | - Chosen plaintext [38] |
| 5 | LBLOCK | [28] | 64 | 80 | Fiestel | - Biclique [48] |
| 6 | LED | [25] | 64 | 64,128 | SPN | - Differential cryptanalysis [39]<br>- Meet-in-the-middle [40] |
| 7 | MIBS | [29] | 64 | 64,80 | Fiestel | - Differential cryptanalysis [49] |
| 8 | PRESENT | [26] | 64 | 80 | SPN | - Cryptanalysis [41-43]<br>- Biclique [44] |
| 9 | PICCOLO | [30] | 64 | 80,128 | Fiestel | - Biclique [44] |
| 10 | SEA | [31] | 96 | 96 | Fiestel | - Cryptanalysis [31] |
| 11 | SIMON | [32] | 32,48,64, 96,128 | 64,72,96,128, 144,192,256 | Fiestel | - Differential Cryptanalysis [50, 51] |
| 12 | SPECK | [32] | 32,48,64, 96,128 | 64,72,96,128, 144,192,256 | ARX | - Differential Cryptanalysis [50, 51] |
| 13 | TWINE | [33] | 64 | 80,128 | Fiestel | - Biclique [52] |

## 3. SIMULATION SETUP AND THE METHODOLOGY FOR EVALUATION

We analyse the number of CPU cycles and RAM, ROM consumption in achieving data security for the secure data aggregation scheme proposed in [9]. The scheme uses the pre distributed keys (proposed in [19]) for the purpose of encryption. We work with the nodes that are constrained regarding:

- Storage
- Communication range (It is assumed that nodes can communicate at least till one hop)
- Battery life

Contiki Cooja is a simulator specifically designed for IoT devices that are having the constraints as described. It is also used as an emulator because the code to be executed by the node is the exact same firmware one may upload to physical nodes [57].

### 3.1. Simulation Setup

This section discusses the simulation results for the considered example scenario. We focus on the data security as it is crucial when designing a data aggregation scheme. Passively acquired data can be used for malicious purpose if confidentiality of data is not taken care of. Our criteria are to measure the number of CPU cycles and RAM, ROM consumption for providing data security. In contiki cooja, there are several options regarding the selection of devices for which one wants to emulate. E.g. Cooja mote, MicaZ mote, CC430 mote, Z1 mote, Sky mote, etc. The comparison between the operating systems TinyOS and Contiki is discussed in [58]. In TinyOS, the application has to be replaced totally when the code is changed. However, the contiki OS is better when it comes to updating the deployed application as it can dynamically replace the changed programs. The protocol proposed in [9, 59] require the code to be updated every time the value of N (total number of nodes in the network) or k (number of consecutive nodes) is changed. Moreover, Contiki supports dynamic loading and unloading of the code and multi-threading. Contiki is an event driven OS and event handlers cannot pre-empt each other. However, interrupts can pre-empt the current running process.

We use Sky motes and Z1 motes of contiki cooja for the purpose of simulation. Sky mote features a 16-bit MSP430 MCU, 10 kB RAM, 48 kB ROM, a cc2420 802.15.4 radio transceiver, an external Flash memory, and temperature, humidity and brightness sensor [17]. Z1 mote has higher configuration and uses MSP430F2617 MCU [18]. The specifications considered for Sky and Z1 motes are as shown in Table 2 and Table 3 respectively.

Table 2. Sky mote Specifications

| | |
|---|---|
| **Flash Memory** | 48 KB |
| **RAM** | 10 KB |
| **Current Consumption** | 20 mA |
| **Operating Voltage** | 3 V |
| **Micro-controller** | MSP430 |

Table 3. Z1 mote Specifications

| | |
|---|---|
| **Flash Memory** | 92 KB |
| **RAM** | 8 KB |
| **Current Consumption** | 19.7 mA |
| **Operating Voltage** | 3 V |
| **Micro-controller** | MSP430F2617 |

## 3.1. Methodology for Evaluation of Ciphers

In this section, we analyse the symmetric key based block ciphers with regard to the number of CPU cycles, energy and memory they consume if applied on Sky and Z1 Motes. The energy is calculated through following steps:

- Add the header file #include “energest.h" in a .c file of the considered cipher.

- To get CPU cycles involved in the different ciphers, add “printf(“energy cpu: %lu", energest_type_time(ENERGEST_TYPE_CPU));" line in PROCESS_THREAD of the .c file.

- To get the power consumption, the formula is: Power(mW) = $\frac{r_x ON}{CPU + LPM} * 20mA * 3V$

Sky and Z1 motes have 20mA current value and 3V voltage. Therefore, CPU cycles received from running the code on Sky or Z1 motes, if multiplied with 60 will give the power consumption in Watts. When this value is multiplied with simulation time, it gives energy consumption in joules. For getting the number of CPU cycles involved in ciphers, we run each cipher separately and follow the steps as discussed. Fig. 4 shows the simulation result, when we run a TWINE cipher on a grid of 20 Sky motes.

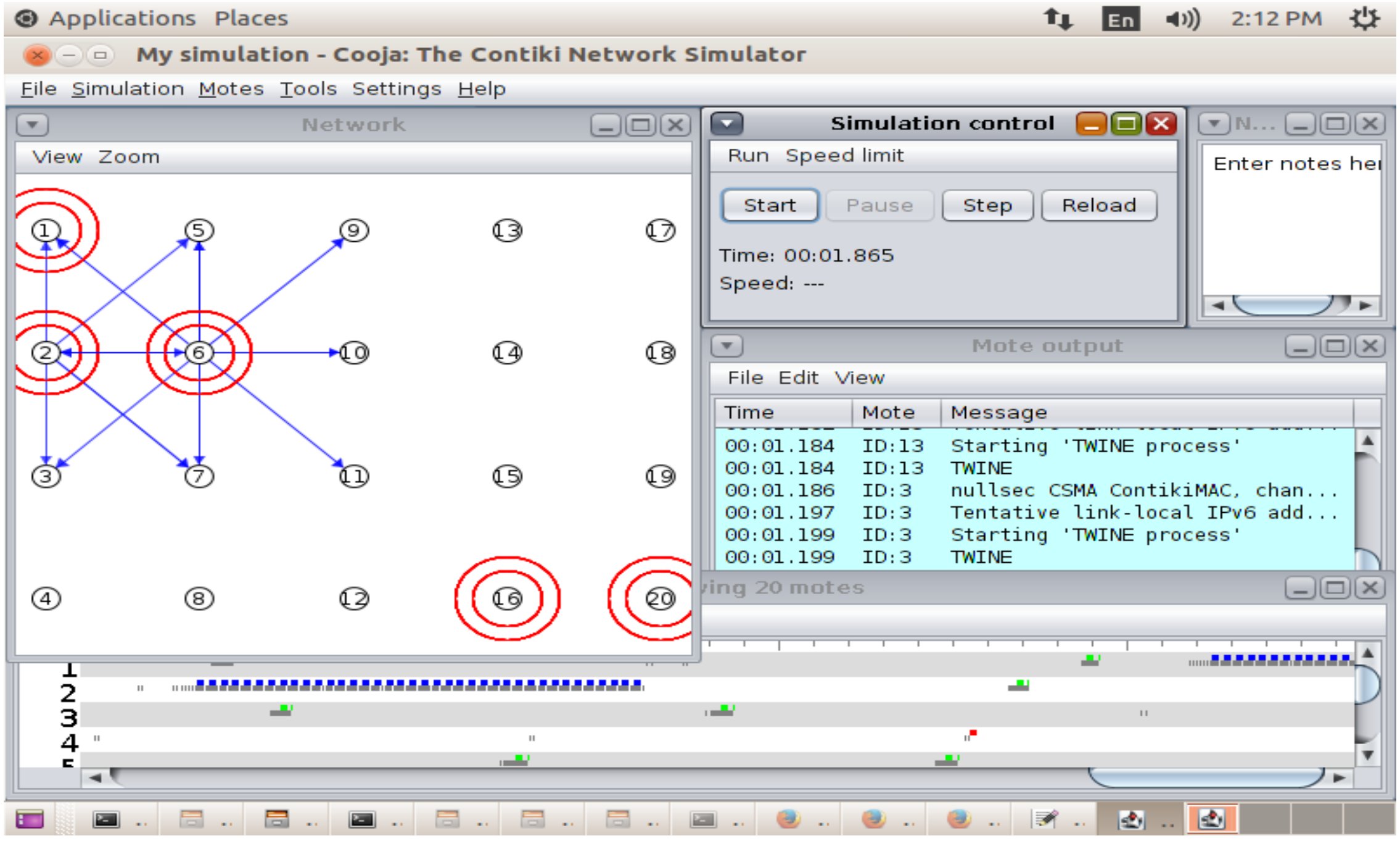

Fig. 4. Simulation of Twine80 Cipher with Sky Mote

In order to obtain the memory (RAM, ROM) consumption, we use “size" command. Fig. 5 shows the use of size command. Here, .text column refers to the ROM consumption by different ciphers in bytes. The .data and .BSS columns show the RAM consumption. We run all the ciphers in the same manner on both Sky and Z1 motes respectively and the results are as shown in Tables 4 and 5.

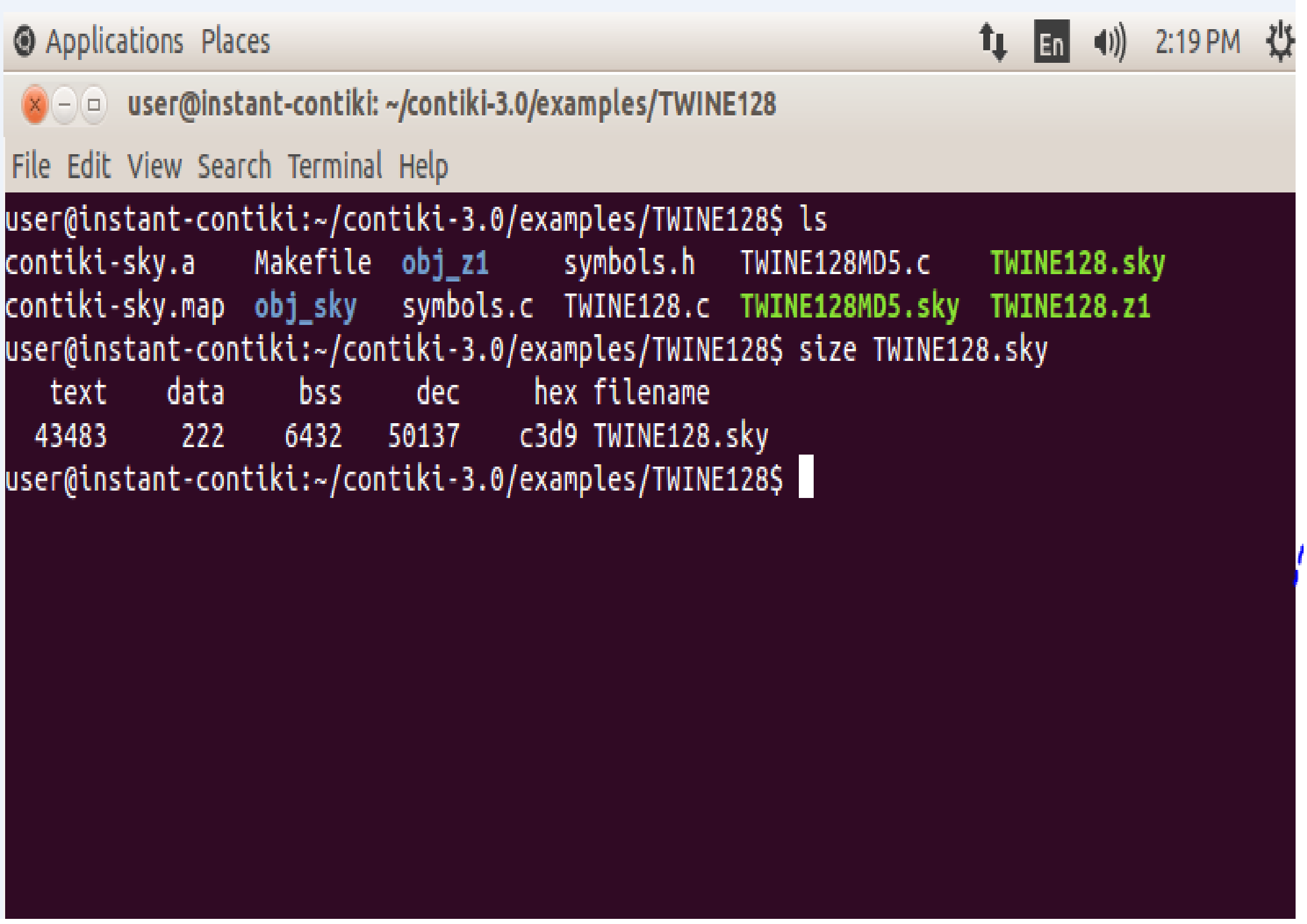

Fig. 5. Size command on Twine Cipher with Sky Mote

## 4. Simulation Results

We can see from the Table 4 that, each cipher requires different amount of CPU cycles and RAM, ROM consumption in the grid network of 20 sky motes. The one that requires maximum number of CPU cycles is LED128 (11117) and the one that requires minimum number of CPU cycles is KLEIN64 (1401). The AES designed by contiki developers specifically for sky motes uses hardware acceleration that helps in reducing number of CPU cycles compared to publicly defined AES (from Table 4, we can see AES (Contiki) requires 1503 whereas AES (Public) requires 1582 number of CPU cycles). SPECK (128 bits block and key size) cipher is designed specifically for resource constrained environments requires 1403 number of CPU cycles.

Table 4. Sky Mote: No. of CPU Cycles and RAM, ROM

| Cipher | CPU Cycles | RAM,ROM (bytes) |
|---|---|---|
| AES (Contiki) | 1503 | 49973 |
| AES (Public) | 1582 | 51329 |
| HIGHT | 1461 | 50093 |
| IDEA | 2375 | 50151 |
| KLEIN64 | 1401 | 50719 |
| KLEIN96 | 1562 | 50763 |
| LBLOCK | 1404 | 50555 |
| LED64 | 7819 | 50149 |
| LED128 | 11117 | 50133 |
| MIBS64 | 1559 | 50381 |
| MIBS80 | 1632 | 50947 |
| PRESENT_Size | 4224 | 51187 |
| PRESENT_Speed | 3715 | 51251 |
| PICCOLO80 | 1487 | 50055 |
| PICCOLO128 | 1512 | 50111 |
| SEA | 1665 | 49923 |
| SIMON128 | 1808 | 50731 |
| SPECK128 | 1403 | 49995 |
| TWINE80 | 1668 | 49943 |
| TWINE128 | 1716 | 50137 |

Table 5. Z1 Mote: No. of CPU Cycles and RAM, ROM

| Cipher | CPU Cycles | RAM,ROM (bytes) |
|---|---|---|
| AES (Contiki) | 633 | 49279 |
| AES (Public) | 285 | 49131 |
| HIGHT | 167 | 48409 |
| IDEA | 643 | 48453 |
| KLEIN64 | 184 | 48995 |
| KLEIN96 | 262 | 49043 |
| LBLOCK | 176 | 48479 |
| LED64 | 2899 | 48499 |
| LED128 | 4342 | 48483 |
| MIBS64 | 242 | 48421 |
| MIBS80 | 277 | 48631 |
| PRESENT_Size | 1300 | 49021 |
| PRESENT_Speed | 896 | 49043 |
| PICCOLO80 | 237 | 48383 |
| PICCOLO128 | 251 | 48451 |
| SEA | 310 | 48319 |
| SIMON128 | 373 | 49079 |
| SPECK128 | 128 | 48347 |
| TWINE80 | 307 | 48333 |
| TWINE128 | 336 | 48531 |

When we run all the ciphers on the Z1 mote, it takes a lesser number of CPU cycles as we can see from Table 5 (AES (pub) on sky mote takes 1582 CPU cycles, whereas on Z1 mote it takes 285 CPU cycles). The AES code provided by contiki cooja developers is specifically designed for Sky mote by using hardware acceleration. Therefore, the RAM consumption of the same is lesser as shown in Table 4. The number of CPU cycles for Z1 motes is always lesser compared to Sky motes, as the configuration of Z1 mote is superior concerning the flash memory that can be used as either RAM or ROM. Therefore, the number of CPU cycles for running different ciphers is lesser for Z1 motes compared to Sky motes as shown in Tables 4 and 5.

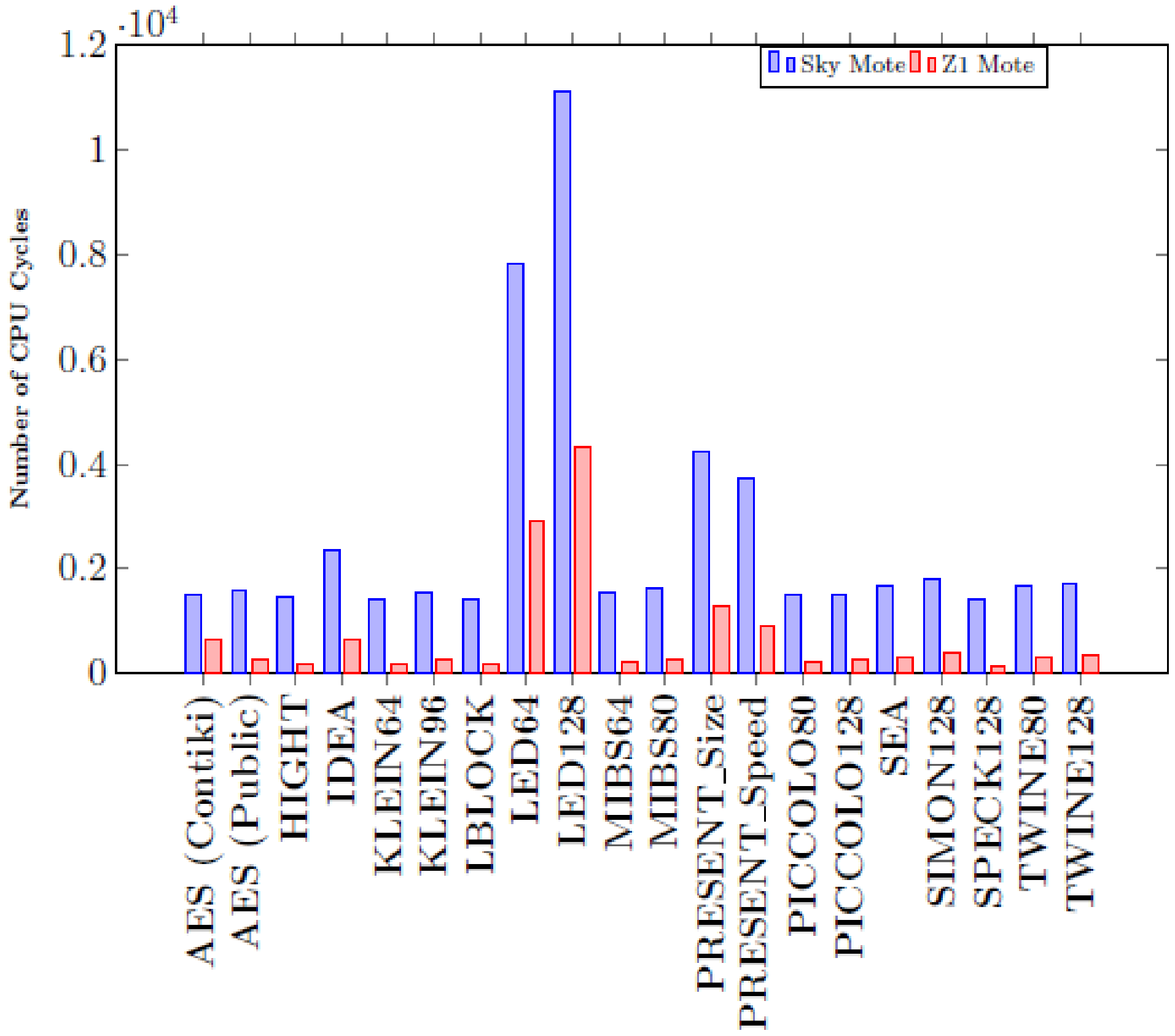

**Fig. 6. CPUCycles**

The comparison of the CPU cycles required by both the motes is as shown in Fig. 6. It shows that Z1 mote takes lesser number of CPU cycles as compared to Sky mote for all the ciphers. The comparison in terms of RAM and ROM consumption of both the motes is as shown in Fig. 7. It shows that Z1 mote takes lesser amount of RAM, ROM consumption as compared to Sky mote in running all the ciphers.

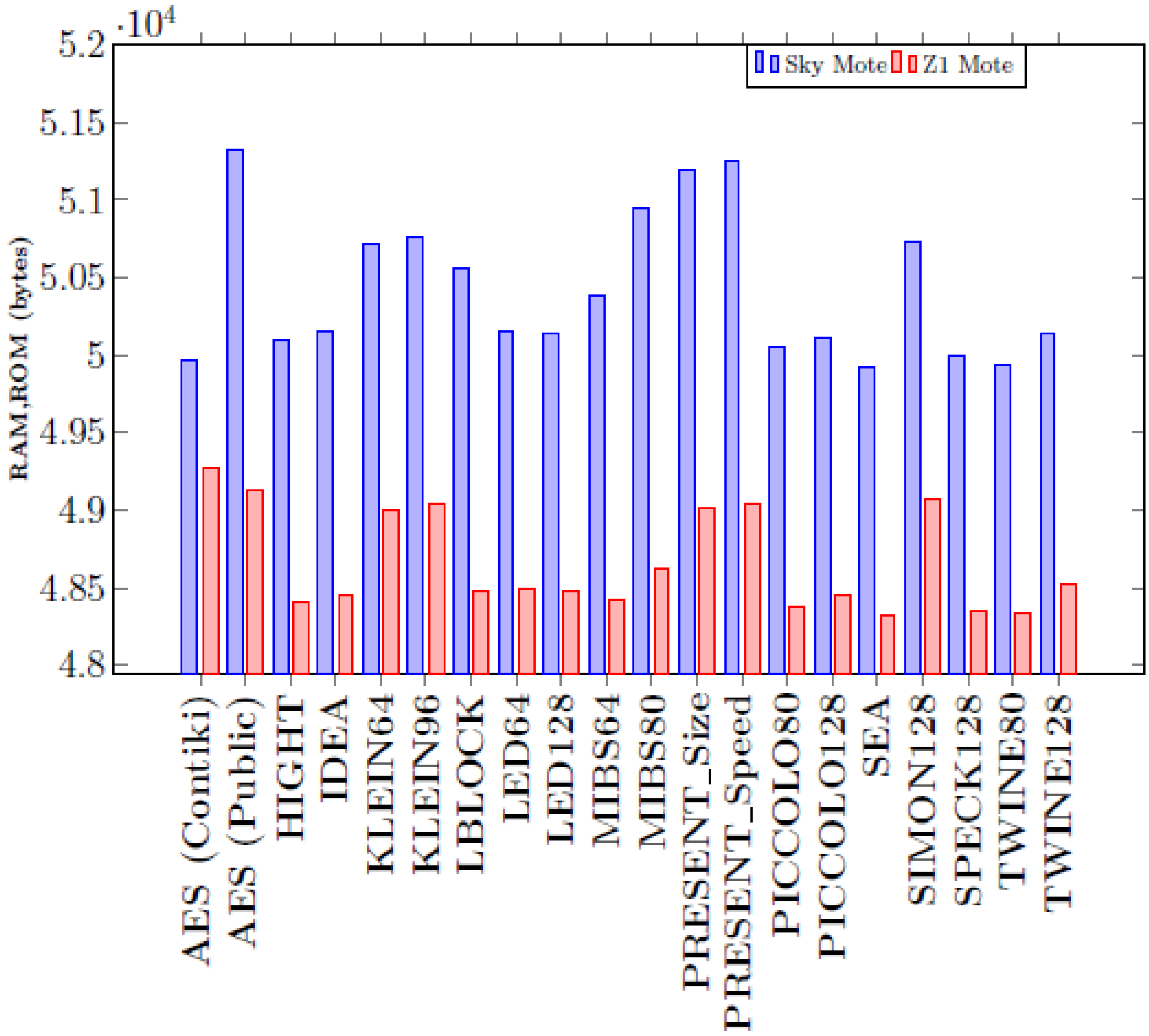

**Fig. 7. RAM,ROM**

## 5. CONCLUSIONS

We analysed symmetric key based block ciphers on two different motes (sky and z1) in contiki cooja. This analysis concerning the number of CPU cycles and RAM, ROM consumption helps in deciding which cipher can be used for the secure data aggregation scheme in different scenarios. In the constrained scenario of sensor nodes, it is better to use lightweight ciphers such as HIGHT; KLEIN; PICCOLO; SIMON; SPECK or TWINE. The hardware accelerated AES cipher by contiki cooja uses minimum number of CPU cycles when we take Sky mote under consideration. Since speed is correlated with energy consumption, SPECK 128/128 is a better choice in energy critical applications as it produces energy efficient solution with an encryption cost of 1403 cycles on Sky mote, or 128 cycles on Z1 mote. We have given a detailed analysis for one of the most significant cryptographic primitives for WSNs: Symmetric Key Block Cipher, by considering the performance parameters, storage and energy for a set of candidate lightweight ciphers. We are working on optimizing the AES cipher regarding speed (reducing the number of CPU cycles) and size (reducing the RAM, ROM consumption). This optimized version of AES will be hardware independent. i.e. it will not depend on the mote under consideration (E.g. sky or z1) and produce the optimal results.

## Authors:

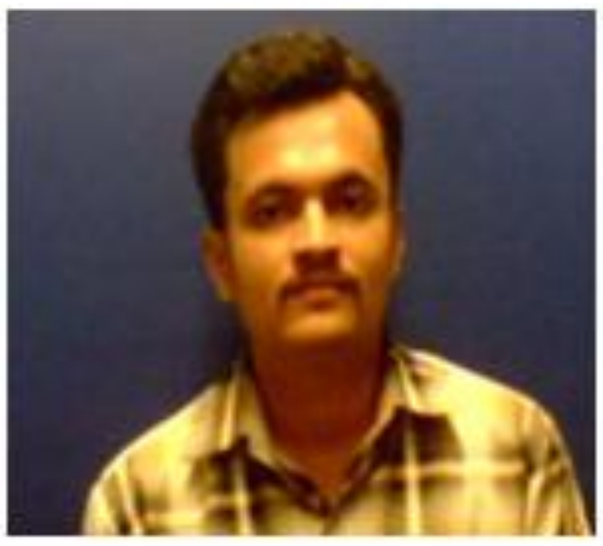

**Kaushal Shah** is a PhD research scholar in Computer Engineering at the Department of Computer Engineering, S. V. National Institute of Technology, Surat, India. He has received his M.E. degree in Computer Science and Engineering from Government Engineering College, Modasa, India. His research interests broadly include Information Security, Wireless Sensor Networks and Protocol designing.

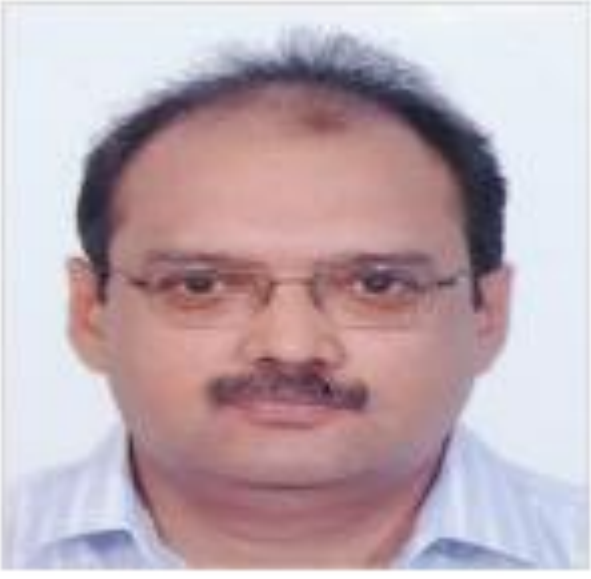

**Devesh Jinwala** has been working as a Professor in Computer Engineering at the Department of Computer Engineering, S. V. National Institute of Technology, Surat, India since 1991. His principal research areas of interest are broadly Security, Cryptography, Algorithms and Software Engineering. Speci_cally his work focuses on Security and Privacy Issues in Resource constrained environments (Wireless Sensor Networks) and Data Mining, Attribute-based Encryption techniques, Requirements Speci_cation, and Ontologies in Software Engineering. He has been/is the Principal Investigator of several sponsored research projects funded by ISRO, GUJCOST, Govt of Gujarat and DiETY-MCIT-Govt of India.